\documentclass[journal]{vgtc}                % final (journal style)
\ifpdf%                                % if we use pdflatex
  \pdfoutput=1\relax                   % create PDFs from pdfLaTeX
  \usepackage{graphicx}                % allow us to embed graphics files
  \DeclareGraphicsExtensions{.pdf,.png,.jpg,.jpeg} % for pdflatex we expect .pdf, .png, or .jpg files
\else%                                 % else we use pure latex
  \usepackage{graphicx}                % allow us to embed graphics files
  \DeclareGraphicsExtensions{.eps}     % for pure latex we expect eps files
\fi%

\graphicspath{{figures/}{pictures/}{images/}{./}} % where to search for the images

\usepackage{microtype}                 % use micro-typography (slightly more compact, better to read)
\PassOptionsToPackage{warn}{textcomp}  % to address font issues with \textrightarrow
\usepackage{textcomp}                  % use better special symbols
\usepackage{mathptmx}                  % use matching math font
\usepackage{times}                     % we use Times as the main font
\usepackage{cite}                      % needed to automatically sort the references
\usepackage{tabu}                      % only used for the table example
\usepackage{booktabs}                  % only used for the table example
\usepackage{setspace}

\usepackage{tikz-cd} 
\usepackage[ruled,vlined]{algorithm2e}
\usepackage{listings}
\usepackage{xcolor}
\usepackage{enumitem}
\usepackage{multirow}
\usepackage{float}

\usepackage{makecell}

\newcommand\realnumberstyle[1]{}
\lstdefinelanguage{JavaScript}{
  keywords={typeof, new, true, false, catch, function, return, null, catch, switch, var, const, let, async, await, if, in, while, do, else, case, break, from},
  ndkeywords={class, export, boolean, throw, implements, import, this},
  sensitive=false,
  comment=[l]{//},
  morecomment=[s]{/*}{*/},
  morestring=[b]',
  morestring=[b]"
}

\lstdefinelanguage{JavaScript}{
  morekeywords=[1]{break, continue, delete, else, for, function, if, in,
    new, return, this, typeof, var, void, while, with},
  morekeywords=[2]{false, null, true, boolean, number, undefined,
    Array, Boolean, Date, Math, Number, String, Object},
  morekeywords=[3]{eval, parseInt, parseFloat, escape, unescape},
  sensitive,
  morecomment=[s]{/*}{*/},
  morecomment=[l]//,
  morecomment=[s]{/**}{*/}, % JavaDoc style comments
  morestring=[b]',
  morestring=[b]"
}[keywords, comments, strings]

\onlineid{1349}

\vgtccategory{Representations \& Interaction}
\vgtcpapertype{Algorithm/Technique}

\title{Semantic Snapping for Guided Multi-View Visualization Design}

\author{Yngve S. Kristiansen, Laura Garrison, and Stefan Bruckner, \textit{Member, IEEE Computer Society}}
\authorfooter{
\item
 The authors are with the Department of Informatics, University of Bergen, Norway. E-mail: ykr088@uib.no, laura.garrison@uib.no, stefan.bruckner@uib.no. 
}

\shortauthortitle{Kristiansen \MakeLowercase{\textit{et al.}}: Semantic Snapping for Guided Multi-View Visualization Design
}
\abstract{Visual information displays are typically composed of multiple visualizations that are used to facilitate an understanding of the underlying data. A common example are dashboards, which are frequently used in domains such as finance, process monitoring and business intelligence. However, users may not be aware of existing guidelines and lack expert design knowledge when composing such multi-view visualizations. 
In this paper, we present semantic snapping, an approach to help non-expert users design effective multi-view visualizations from sets of pre-existing views. 
When a particular view is placed on a canvas, it is ``aligned'' with the remaining views--not with respect to its geometric layout, but based on aspects of the visual encoding itself, such as how data dimensions are mapped to channels. Our method uses an on-the-fly procedure to detect and suggest resolutions for conflicting, misleading, or ambiguous designs, as well as to provide suggestions for alternative presentations. With this approach, users can be guided to avoid common pitfalls encountered when composing visualizations. Our provided examples and case studies demonstrate the usefulness and validity of our approach.%
} % end of abstract


\CCScatlist{ % not used in journal version
 \CCScat{K.6.1}{Management of Computing and Information Systems}%
{Project and People Management}{Life Cycle};
 \CCScat{K.7.m}{The Computing Profession}{Miscellaneous}{Ethics}
}

\teaser{
  \centering
  \includegraphics[width=0.955\linewidth]{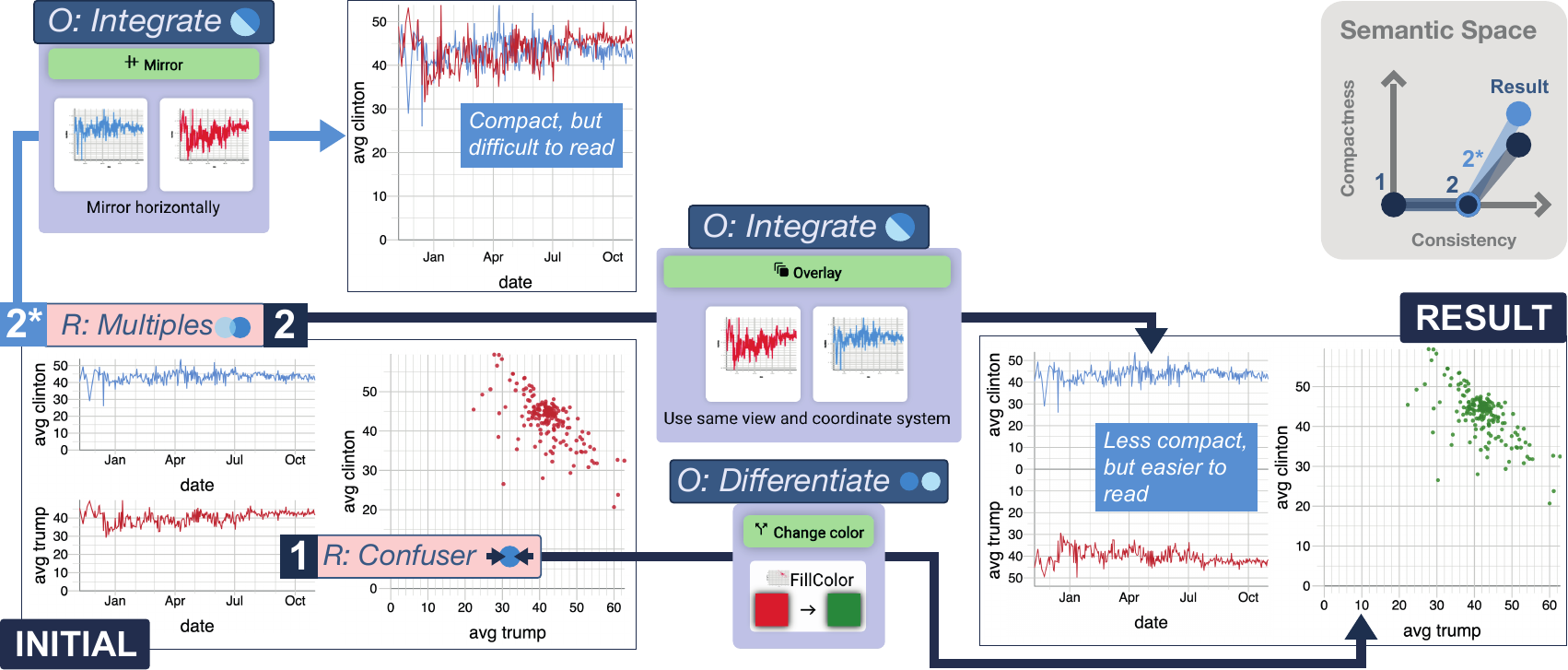}
  \caption{Semantic snapping allows the user to perform iterative operations to improve the compactness and consistency of a multi-view visualization. Underlying algebraic rules called \textit{relations} define the available \textit{operations} for each iteration. In this example showing the 2016 US Election poll percentages and pollsters, from our initial composition we (1) identify a \textit{confuser} relation between the bottom left and rightmost views showing the same color (red). We \textit{differentiate} these views by selecting green as the fill color for the scatter plot. We next identify a \textit{multiples} relation in the two left views. We resolve this through one of two \textit{integration} operations. (2*) \textit{Overlay} produces an unsatisfactory result, so we revert and (2) perform a \textit{mirroring} operation to arrive at our resulting composition. The semantic map to the right illustrates our path through semantic space.\\}
  \label{fig:teaser}
}

\vgtcinsertpkg

\begin{document}

%% The ``\maketitle'' command must be the first command after the
%% ``\begin{document}'' command. It prepares and prints the title block.

%% the only exception to this rule is the \firstsection command
\firstsection{Introduction}

\maketitle

%% \section{Introduction} %for journal use above \firstsection{..} instead

% What (broad)

%In recent times topics such as data science are becoming more prevalent, and the amount of data to analyze is increasing. 
Multi-view visualizations are frequently utilized to present and analyze data. Dashboards, for example, are commonly employed for monitoring and related tasks in a wide variety of fields. Popular visualization systems like  Tableau \cite{stolte2002polaris} and PowerBI provide galleries of carefully crafted templates in order to enable the quick and easy generation of such visualizations. However, when non-expert users would like to extend, modify, or customize such a multi-view visualization, they may easily fall prey to a number of pitfalls that can result in potentially misleading or otherwise problematic results. Expert knowledge to guide such tasks is mostly available in the form of guidelines from the visualization literature, which are not readily accessible to novice users. Common examples include Qu and Hullman's constraints, C1 (the same data should be shown the same way) and C2 (different data should be shown in different ways)~\cite{QuEvaluatingVisSets2016}.
%Another problem a multi-view visualization designer may face is when the design uses more screen space than intended. In these cases, it is preferable to communicate the same information with less screen space, i.e., more compactly.
In this paper, we present a method that detects and helps users to resolve such potential problems in multi-view visualization design in a semi-automatic and guided fashion.

Suppose a user of a visualization system wants to create or extend a multi-view visualization from a set of pre-existing charts that are individually well-designed (e.g., based on a gallery). If the user wishes to use these visualizations in combination (e.g., in a dashboard), there are non-obvious design opportunities and pitfalls. Views may show the same data in different ways, or different data in the same way. A single view may be highly informative and take up a modest amount of screen space. However, when used in combination with other views, they may show overlapping information, or use too much screen space. These issues can be remedied by showing the same data with fewer views, i.e., making the overall design more \textit{compact}.
%but show overlapping information, unnecessarily taking up more screen space than what is available. 
%Other views may be combined into one view that shows exactly the same data. 
A multi-view visualization can be made more compact, or less conflicting, by manually redesigning and tweaking single views. However, manually detecting and resolving conflicts, and coming up with alternate representations of views, is cumbersome, error-prone and time-consuming. Our method lets the user perform this process via high-level design-altering operations.

Our approach uses a semantic space with two axes that represent the degree of consistency and compactness of a multi-view visualization. We examine \textit{relations} between views to identify opportunities to improve the overall visualization with respect to these criteria. We then provide the user with a set of operations to realize the corresponding changes. For instance, two views may employ the same color map for different quantities. In such a case, our approach may suggest to \textit{differentiate} the two views by modifying one of the mappings to increase the overall consistency. Likewise, when the same data are shown differently in multiple views, our method suggests different ways to \textit{homogenize} them. In other cases it may be possible to \textit{integrate} multiple views in order to improve the compactness of the visualization. 

The contributions of our work can be summarized as follows. Based on a synthesis of existing guidelines from the literature, we present a novel approach for identifying and applying potential improvements of multi-view visualizations. We use predicate logic to represent relations between individual views and propose operations to improve the consistency and compactness of the underlying visualization based on the identified relations. Furthermore, we propose a simple workflow and user interface for presenting and selecting the suggested operations.

%-----------------------------------------------
\section{Related Work}
%TODO need to add something high level here about the work and where it's drawing from before launching in to indvidual related work bits 

\textbf{Visualization design and measures. } % single view
%The visualization design process is often not only guided by objective criteria, but also by the designer's personal preferences. Standards such as guidelines and effectiveness measures aids objective and computational reasoning about visualizations. Much work has been done to investigate exactly what factors contribute to a visualization's effectiveness of lack thereof. 
%
%In order to reason about the effectiveness of visualizations, it must also be possible to reason about visualizations. Bertin's Semiology of Graphics \cite{Bertin:1983:SG:1095597} and Wilkinson's Grammar of Graphics \cite{Wilkinson:2005:GG:1088896} enable for reasoning about graphics. Munzner \cite{Munzner:VisAnalysisAndDesign} introduced the term \textit{channel} as a way to control the appearance of graphical marks. In our method, we use channels to describe components of a visual encoding, but divide them up into three components for fine-grained reasoning about across across multiple views. 
%
Bertin's Semiology of Graphics~\cite{Bertin:1983:SG:1095597} and Wilkinson's Grammar of Graphics \cite{Wilkinson:2005:GG:1088896} were two of the early influential works focusing on formal aspects of reasoning about the effectiveness of visualizations. Munzner~\cite{Munzner:VisAnalysisAndDesign} later consolidated and refined existing concepts and terminology, leading to a comprehensive framework for thinking about visualization in terms of principles and design choices.
Bolte \& Bruckner~\cite{BolteMeasuresVisSpace2020} survey measures focusing on different aspects of the visualization process: perceptual characteristics, task-oriented quality measures, structure-oriented measures, and meta-perceptual processes. Perceptual characteristics such as Cleveland \& McGill's experiments on graphical perception~\cite{cleveland1984graphical} are based on human performance in elementary tasks such as comparing positions on a common scale. Other measures express desirable relationships between the data and its visual representations. For example, Tufte's data-to-ink ratio~\cite{tufte1985visual} describes the proportion of pixels used to represent data versus the total amount of available pixels. Furthermore, Correll et al.~\cite{correll2018looks} address the issue that designs may appear to be showing the data completely, while hiding important details. They propose actions to remedy discovered vulnerabilities for different chart types. Behrisch et al.~\cite{behrisch2018quality} categorized different quality measures from around 250 papers. Most of these measures were specific to a certain combination of underlying data, task and visualization technique.

Through literature review, Zhu~\cite{zhu2007measuring} points out why existing definitions of visualization effectiveness are often incomplete: they usually take either a data-centric, or task-centric view on what an effective visualization is. 
%He mentions that much research takes either a data-centric or task-centric view on what an effective visualization is. Their characterization of the data-centric perspective is largely in line with Bolte and Bruckner's category of structure-oriented measures.
%Even though these highly useful effectiveness metrics are not explicitly a part of our approach, we believe they could be used and also adapted for evaluating multi-view visualizations. Through simulations and a crowdsourced study, they detect different vulnerabilities for different charts, and provide recommendations to improve the reliability of each chart type. We do not do take into account vulnerabilities of chart types, but believe the findings of Correll could be highly useful if applied, and possibly adapted for, multi-view visualizations.
Data-centric effectiveness measures deal with how accurately a visualization is showing its underlying data. An example of a data-centric framework for measuring visualization effectiveness is Kindlmann and Scheidegger's algebraic framework~\cite{kindlmann2014algebraic}. By considering symmetries between changes in data space and resulting changes in visualization space, they describe three principles that should ideally be true for any data-to-visualization mapping: unambiguous data depiction, representation invariance, and visualization-data correspondence. We draw inspiration from this model and adopt a similar line of reasoning in the context of multi-view visualizations. 
Based in part on the concept of algebraic visualization design, McNutt and Kindlmann~\cite{mcnutt2018linting} present a linting mechanism for the process of designing a chart. Their linting is realized as a Python library that evaluates visualizations created with matplotlib, and returns a list of rules that are violated. While our work is based on similar fundamental considerations, we focus on multi-view visualizations and expose potential revisions through a user interface. 

Many approaches take into account both data and tasks. Cantu et al.~\cite{cantu2017identifying} outline an approach to identify relationships between visualization challenges and representation components (e.g., data transformations, filtering techniques, visual variables). They argue that these relationships can further our understanding of the mechanisms behind visualization components, which could eventually be used to build visualization recommendation tools. Silva et al.~\cite{silva2007there} survey work done on using different color scales in visualization, with a focus on desired properties and guidelines for choosing the right colors. They highlight that it is important to consider factors such as the type of data, type of visualization, type of task, and audience.
As pointed out by Zhu~\cite{zhu2007measuring}, there are multiple disjoint, sometimes conflicting sets of guidelines and measures. Efforts have been made to facilitate convergence and understanding between different viewpoints. Diehl et al.~\cite{diehl2018visguides} initiated the VisGuides forum both to facilitate collection and discussion of visualization guidelines, and knowledge about visualization in general. Engelke et al.~\cite{engelke2018visupply} highlight that there is a gap between the communities who propose visualization guidelines, and those who need them. They provide a conceptual model called VISupply that highlights problems and opportunities with how guidelines are currently ``shipped'' to non-experts. 

% Design process guided by standards & measures
% --> description of measures for effective vis
% --> viszrec as a "result" of all these
% and personal preferences --> vis authoring tools

\textbf{Authoring tools and visualization recommendation. }
Visualization authoring tools help users creatively express a wide range of individual charts. While these systems have much design freedom, they also rely on the expertise of the user. Zhu et al.~\cite{zhu2020survey} survey different tools for automatically generating infographics and visualization recommendations.
%He classifies them into three categories: knowledge-based, data-driven and hybrid visualization design tools. 
Examples of systems that mostly focus on authoring and design flexibility include Charticulator~\cite{2019-charticulator}, Lyra~\cite{2014-lyra}, iVisDesigner~\cite{ivis-designer}, Data Illustrator~\cite{data-illustrator-liu}, Data Driven Guides~\cite{data-driven-guides-kim}. These systems all use varying underlying frameworks for representing visualizations. We provide a set of relations and operations specified at a high enough level so that they can be expressed in terms of most individual frameworks. 
%We do not provide an underlying framework or designer for individual charts, but a conceptual set of rules specified at a high enough level so that they can be expressed in terms of most individual frameworks. 

% Redundant I think, can be mentioned after next block
%Similar to this work, we aim to provide users with a set of guidelines, but we do so indirectly and in a semi-automatic guided fashion by providing the user with operations to revise a multi-view visualization design. 
Several efforts have been made to make expert knowledge available through software. Among them, visualization recommendation systems can potentially take into account expert knowledge to steer which revised designs are presented to the user. MacKinlay's APT (A Presentation Tool)~\cite{mackinlay1986automating} was among the first of these systems. He used a composition algebra for designing visualizations, and evaluated their effectiveness in accordance with Cleveland \& McGill's effectiveness metrics~\cite{cleveland1984graphical}. Wongsuphasawat et al. proposed CompassQL~\cite{2016-compassql} as a general language for querying over the space of visualizations, to be used in visualization recommender systems. Voyager~\cite{wongsuphasawat2015voyager} allows for exploring data via automatically generated visualizations. 
With Voyager 2~\cite{wongsuphasawat2017voyager2}, the user is able to partially specify what a view should show by using wildcards and also see automatically-generated charts showing data related to the existing views. % Partial specifications are useful for reasoning and used in semsnap
Data2Vis~\cite{dibia2019data2vis} is a trainable neural translation model for automatically generating visualizations from datasets. It is powered by formulating visualization generation as a language translation problem, where data specifications are mapped to Vega-Lite specifications~\cite{2017-vega-lite}. 
Grammel et al.~\cite{grammel2010information} explore how novices construct visualizations. Their findings suggest the need for a tool that supports iterative refinements, and explanations that help with learning. Our method shares a similar line of thought by enabling incremental refinement of a multi-view visualization.
Show Me~\cite{mackinlay2007show} is a set of user interface commands that provide a way to display an additional data attribute within a view, as well as high-level commands for building views for multiple fields. 
Draco~\cite{2019-draco} makes visualization design guidelines available for a wider audience by formalizing the knowledge into precise constraints, which can then be used and accessed in an Answer Set Programming environment. They model single visualizations as sets of logical facts, and represent design guidelines as hard and soft constraints over these facts. Dziban~\cite{lin2020dziban} further extends Draco with anchoring mechanisms to help drive specification queries with increased user agency.
These works all represent different ways of representing and reasoning about visualizations. 
Our method differs from these approaches in that it focuses in the incremental refinement of multi-view visualizations.

\textbf{Multi-view visualization design. } %
One of the most common use-cases of multi-view visualizations are dashboards. 
% A little bit about dashboards
Sarikaya et al.~\cite{sarikaya2018talkdashboards} construct a design space of dashboards, by analyzing multiple examples of dashboards found ``in the wild.'' 
%They also contend that there are design principles that apply to all kinds of dashboards, among them, consistency between views. 
QualDash~\cite{elshehaly2020qualdash} is a task-oriented dashboard generation engine that enables the mapping of specific user task sequences in healthcare quality improvement to a view composition.
%
% Some about multi-window / multi-device
For dashboards and multi-view visualizations in general, multiple views must be laid out on a single screen, or even multiple screens. PanoramicData~\cite{zgraggen2014} is a visual analysis tool using a canvas metaphor to explore and combine data views. We use a similar metaphor in our approach, although we focus on semantics rather than filtering and linking the views. 
Vistribute~\cite{horak2019vistribute} is a framework that automatically distributes visualizations and user interface components among multiple heterogenous devices.
Scout~\cite{swearngin2020scout} is a system that helps interface designers to create layouts by using high-level constraints based on design concepts such as semantic structure, emphasis and order. While our approach currently does not address layout, we believe that our method could be combined with similar approaches to also take into account layout considerations.
%\textit{Semantic Snapping} uses a similar approach to Scout in that it helps to generate a more diverse range of designs.

% Now techniques for composing / comparing views etc, small multiples, etc etc
Composed views such as small multiples \cite{Tufte:1990:EI:78223} allow for comparing visualizations. Gleicher et al.~\cite{gleicher2011visual} provide a general taxonomy of visual designs for comparing visualizations, with three categories: juxtaposition, superposition and explicit representation of relationships. 
Elzen and van Wijk~\cite{van2013small} leverage small multiples so that they are not only informative, but also helpful for the data exploration process itself. 
%\textit{Semantic Snapping} is similar to visualization approaches in that it provides alternative designs. Our method stands out from these approaches in that each alternate design specifically resolves a detected problem in the existing design. 
Through a series of graphical perception experiments, Ondov et al.~\cite{ondov2018face} investigated which compositions of multiple charts are the most effective for different tasks.
From 360 images of multi-view visualizations collected from IEEE VIS, EuroVis and PacificVis publications from 2011 to 2019, Chen et al. \cite{chen2020composition} identify common multi-view visualization practices, including typical view layouts, view types, and correlations between view types and layouts. The patterns found among these views are made available through a multi-view visualization recommendation system, allowing users to interactively browse different designs. 
We draw inspiration from these approaches by enabling the transformation of, for example, two bar charts into an item-wise grouped or chart-wise juxtaposed mirrored bar chart in order to increase the compactness of the overall visualization, as in Figure~\ref{semsnap:fig:ui_examples}.

Conventional snapping creates a ``gravity field" around geometric objects, making it easier to place them together in certain ways. Hudson \cite{hudson1990adaptivesemanticsnapping} introduced the notion of \textit{semantic snapping} as an interaction technique for geometrically snapping objects together only if the objects are specified to be semantically related. Our work is a continuation of this basic concept, extending it to the scenario of multi-view visualization design and focusing on the semantic rather than geometric aspects.
Shadoan \& Weaver~\cite{shadoan2013} explore semantic relations in multi-view visualizations using a hypergraph querying system. While such queries are constructed similarly to \textit{relations} in our approach, the former are driven through cross-filtering on attribute relationship graphs, while ours draws from rules heavily inspired by Kosslyn's principles~\cite{kosslyn1989understanding} and Kindlmann and Scheidegger's algebraic framework~\cite{kindlmann2014algebraic}. The latter framework has been used to identify effective visualization types for certain user tasks, e.g., table cartograms~\cite{mcnutt2021table}. 
Kim et al. characterize responsive visualization strategies via their targets, i.e., element(s) of a design that change, and actions, i.e., how element(s) are changed~\cite{kim2021design}. This semantics-based characterization parallels our notion of \textit{relations} and \textit{operations}, although the underlying models differ. 

Qu and Hullman~\cite{QuEvaluatingVisSets2016} discuss how to operationalize Kosslyn's principles~\cite{kosslyn1989understanding} with the two following constraints: C1 (encode the same data in the same way), and C2 (encode different data in different ways). 
These two constraints are further detailed by specifying lower-level constraints on encodings across two views.
In a later paper~\cite{QuKeepingMultipleViewsConsistent2018}, they found through a Wizard-of-Oz study that Tableau users unknowingly, and with some exceptions, respected their constraints C1 and C2. They found that study participants were positive to having a consistency checker tool to surface such warnings. Similarly to Qu and Hullman, we operationalize the principles C1 and C2 on an encoding-level, but we do so by using a model inspired by Kindlmann and Scheidegger's algebraic framework~\cite{kindlmann2014algebraic}. Furthermore, we present a practical realization of this concept that both shows how to identify potentially problematic relations and introduces a set of concrete operations to address the relations, i.e., remove the relation itself or a problem caused by the relation.
% TMI? 
%The algebraic model's \textit{hallucinator} and \textit{confuser} is semantically equivalent to violations of Qu and Hullman's C1 and C2, and chose to use the terms from the algebraic model, since that is closest to our realization. Furthermore, we provide a realization of our model in the form of a semi-automatic guided design tool, where the user can access operations that alter the design -- each resolving a detected relation.

%-----------------------------------------------
\section{Semantic Snapping Model}
% What is it
% Why do we do it?
%EXAMPLE! Consider a novice user who has already defined a set of individual charts, and wants to combine them into a multi-view visualization such as a dashboard. While the charts individually may be well-designed, new considerations must be taken when combining them.
% How (Semantic Axes)
 
% Semantic Axes to relations
% Relations to operations
% Short TLDR of method + overview
%The process of \textit{Semantic Snapping} begins with an existing multi-view design. With this existing design, our method identifies existing relations between all sets of views. Each relation identifies a potential problem, which can be resolved by an operation.% The cycle of finding relations to infer available operations is repeated every time the design is altered. 

\begin{figure}[tb]
    \centering
    \includegraphics[width=0.98\linewidth]{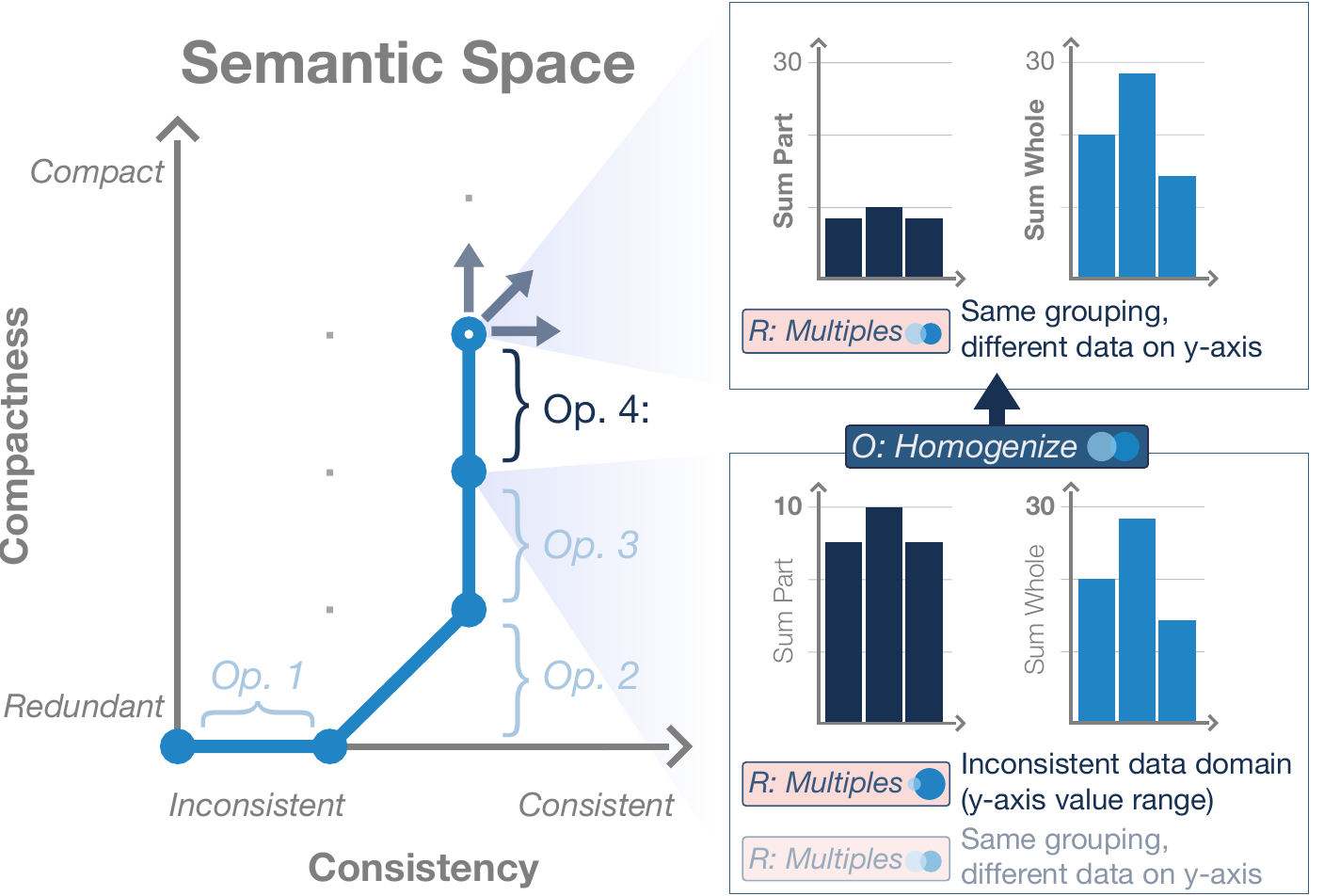}
     %\vspace{-0.1cm}
    \caption{Conceptual figure showing the semantic space with relations and operations for our semantic model. Operation 4 displays the \textit{homogenize} operation which is available as a result of a \textit{multiples} relation (the two axis scales are different, but should be the same if the underlying data represents the same quantity). %data of the two charts are semantically the same, but the visual outputs are different due to them using different data domains. The homogenize operation makes the data domains the same. 
    }
    \label{fig:concept-figflow}
\end{figure}

%\textit{Semantic Snapping} is the process of incrementally modifying a multi-view visualization by aligning its individual views with respect to semantic, rather than their geometric, attributes. It is an incremental workflow whereby semantic redundancies and inconsistencies are identified and resolved through high-level changes. Each high-level change has an understandable and predictable effect. This approach allows the user to explore high-level design decisions without needing to manually piece together low-level modifications. 
Semantic snapping is the process of incrementally modifying a multi-view visualization by aligning its individual views with respect to their semantic, rather than their geometric, attributes. The main underpinning of our method is that a multi-view visualization, and its potential revisions, can be placed into a semantic space with two dimensions representing the degree of compactness and degree of consistency. We provide a conceptual overview of this semantic space in Figure~\ref{fig:concept-figflow}. In other words, a potential revision of a design is either more or less compact, or more or less consistent, than the original design. Our method identifies these potential revisions, and presents them to the user as operations. %in a predictable and understandable way. 
By executing these operations, the user is able to intuitively navigate the semantic space of revised designs. 
%Consider the following scenario: A user has defined a set of individual visualizations and uses them to construct a multi-view visualization, e.g., a dashboard. While visualizations individually may be well-designed, new considerations must be made when they are used in combination. The overall design may become inconsistent or redundant.
%If such problems arise, it is typically up to the user to resolve them. For example, suppose two visualizations are showing similar, but not exactly the same data, and the user believes that one visualization could show the data shown by both of these views. Typically the user would create a new visualization showing all the data from scratch, and then insert it into the design. 
%The problem with this approach is that the onus is on the user to identify inconsistencies and redundancies in their visualization and determine how to best solve them. Furthermore, the user has to manually express the solution through a series of low-level modifications. This may be simple for a practiced visualization designer, but may prove challenging and prone to trial-and-error for a non-expert user. 

Achieving high-level goals by piecing together low-level modifications can be tedious, especially for novice users. In other software, such as word processors, semi-automatic tools help with this workload by highlighting errors, and suggesting corrections to these errors. Previous approaches, such as McNutt and Kindlmann's linting mechanism \cite{mcnutt2018linting}, have already explored this direction by providing functionality akin to a spell checker for a \textit{single} visualization. In contrast, semantic snapping can be seen as more similar to a grammar checker, since it focuses on relationships \textit{between} visualizations, just as a grammar checker analyzes relationships among words or phrases. Errors or potential errors represent detected inconsistencies or redundancies between views, and error corrections are represented as suggested operations to revise the composition of views.

Our method identifies existing and potential semantic inconsistencies and redundancies, so-called \textit{relations}, between single views. Each relation identifies a potential problem which can be resolved by an \textit{operation}. Thus, each operation is a high-level modification to the overall design. It is necessary to have the user involved in each modification to the design, since consistency and compactness are sometimes traded off for other design considerations \cite{QuKeepingMultipleViewsConsistent2018}. The cycle of finding relations to infer available operations is repeated every time the design is altered. 

\subsection{Semantic Space} % Discuss here how the frame of a semantic space is established

\begin{figure}[tb]
    \centering
    \vspace{-0.1cm}
    \includegraphics[width=\linewidth]{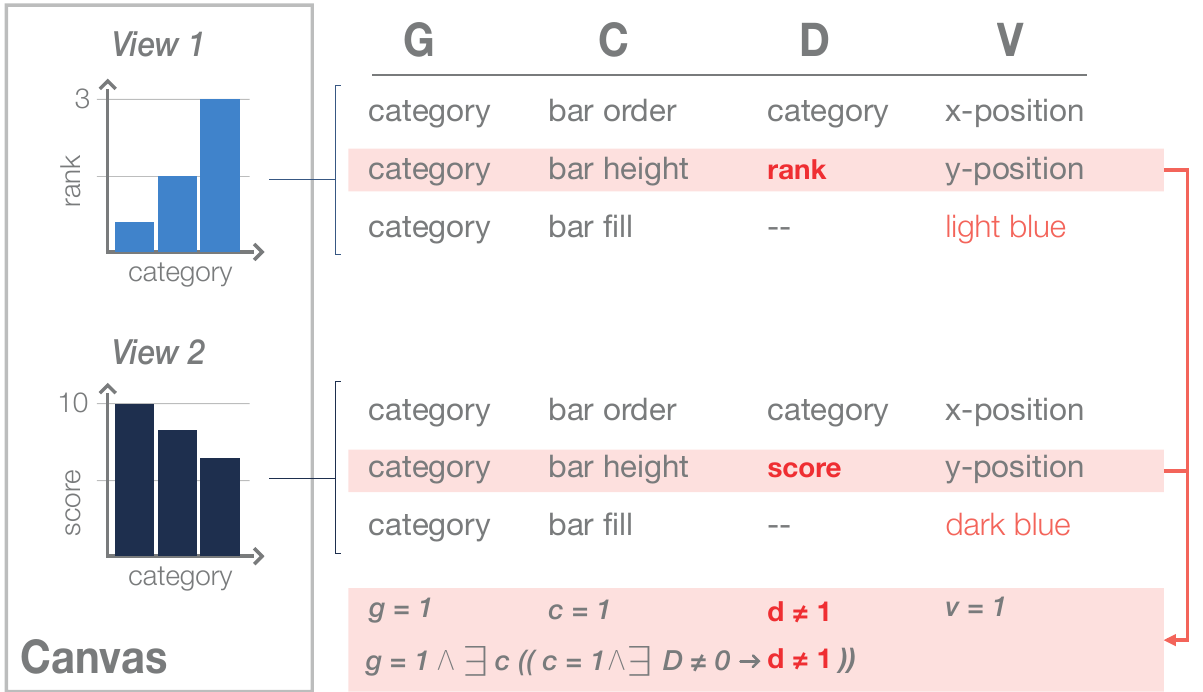}
    \caption{Here we see two views, and the (G, C, D, V) tuples corresponding to the three channels representing the y-axis (bar height), x-axis (bar order), and fill color. The comparison of the two highlighted tuples are what identifies these two views as \textit{multiples (1) same grouping}. }
    \label{fig:semsnap:relations-conceptual}
\end{figure}

\begin{table*}[tb]
    \centering
    \begin{tabular}{l l|l|m{50pt}|l | m{50pt}}
         & Relation & Specification & Illustration & Possible Operations & Illustration \\
         \midrule
         (a) & Full redundancy & $g=1 \land \forall{c}(c=1 \rightarrow d=1)$ &{ \hfil\includegraphics[width=50pt]{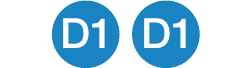}} & Delete one & 
            \hfil\includegraphics[width=50pt]{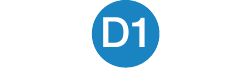}
         \\
         %& & & & & \\ 
         (b) & Partial redundancy & $g=1 \land \forall{c}((c=1 \land d \neq 1) \rightarrow \exists!{D} = 0)$ & \hfil\includegraphics[width=50pt]{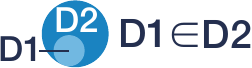} & Integrate, or delete D1 view  & 
            \hfil\includegraphics[width=50pt]{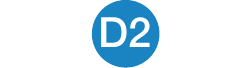}\\  
         %& & & & & \\ 
         %\hline
         (c) & \makecell[l]{Multiples (1) \\ \small{\vspace*{-0.15cm} same grouping}} & $g=1 \land \exists{c}((c=1 \land \exists{D} \neq 0) \rightarrow d \neq 1)$ & 
         \hfil\includegraphics[width=50pt]{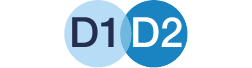}
          & Integrate or homogenize &
         \hfil\includegraphics[width=50pt]{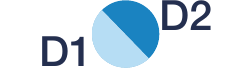}
          \\
          %& & & & & \\    
         (d) & \makecell[l]{Multiples (2) \\ \small{\vspace*{-0.15cm} same data}} & $g \neq 1 \land \exists{c}((c = 1 \land \exists{D} \neq 0) \rightarrow d = 1)$ & \hfil\includegraphics[width=50pt]{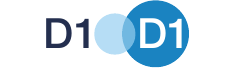} & Homogenize & \hfil\hfil\includegraphics[width=50pt]{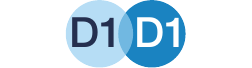} \\ %& & & & & \\
          %& & & & & \\
          %& & & & & \\
         %\hline
         (e) & Hallucinator & $g=1 \land \exists{c}(c=1 \land d=1 \land \exists{D} \neq 0 \land v \neq 1)$ & \hfil\includegraphics[width=50pt]{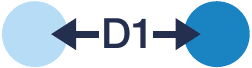} & Homogenize & \hfil\includegraphics[width=50pt]{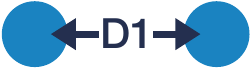} \\
         %& & & & & \\
         %\hline
         (f) & Confuser & $\exists{c}(c=1 \land d \neq 1 \land v=1)$ &  \hfil\includegraphics[width=50pt]{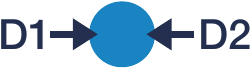} & Differentiate &
         {\hfil\includegraphics[width=50pt]{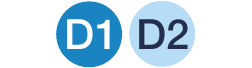}}
         \\
         %\bottomrule
    \end{tabular}
    \vspace{0.5\baselineskip}
    \caption{All relations specified in terms of our model. The lower case letters: $g$, $c$, $d$, and $v$ represent equalities ($1$) or inequalities ($0$) between chart groupings, channels, data mappings and visual outputs, and the specifications are predicate logic expressions operating primarily on these (lower case) equalities or inequalities. The uniqueness quantifier on $\exists{!}D$ indicates that there exists exactly one data mapping that satisfies a certain condition, for example being unmapped ($= 0$) for (b). 
    %Note how the $d \eq 1$ clause in the \textit{multiples(2)} and \textit{hallucinator} relations evaluates to true if the data mappings are different, but represent the same semantic quantity.
    %Note how the \textit{multiples(2)} and \textit{hallucinator} relation has two instances, where the one with $d \neq 1$ is a potential relation, depending on whether the data are semantically equal. 
    For the \textit{partial redundancy} relation, $D1 \in D2$ signifies that all data shown by one view ($D1$) is also shown by the other view ($D2$). }
    \label{tab:semsnap_relations}
\end{table*}

We begin with defining the terms that comprise our semantic space. A \textit{canvas} is composed of multiple views of a single tabular dataset, where each individual view displays a single chart of a certain type (for example a bar chart, scatter plot, line chart, etc.).
A \textit{chart grouping} is a data dimension by which a chart is grouped, similar to SQL's \lstinline[language=SQL]{GROUP BY} command. For example, a bar chart grouped by country will have one bar for each distinct country in the dataset. Each chart has a set of \textit{channels}, which may or may not be mapped to data.
Data shown by a channel is denoted as a \textit{data mapping}, which may also be empty (indicating an unmapped channel). For example, consider the fill color channel in a scatter plot, which can optionally be mapped to data to display an additional quantitative data attribute for each mark. 
When a channel has a data mapping, the data are scaled from a \textit{data domain} to a resulting \textit{visual output} which directly or indirectly affects the appearance of the chart. The data domain denotes the minimum and maximum value of a certain attribute, or attribute aggregate, and is used as an input for the scale from data to a resulting visual output. When groupings differ, sampled data domains may also differ. Furthermore, domains may be different due to custom configuration of individual views. While there are different ways to arrange and transform tabular data, we limit the scope of chart groupings and channel mappings to single data dimensions. 
%We also consider the \textit{visual output} of a channel for reasoning about multiple views. The visual output directly or indirectly affects the appearance of the shapes of the chart, and is affected by both the output retinal range, and data domain in use. For example, if the fill color channel has a certain data mapping, the visual output may be any color scheme (for example reds or greens).
Summarized, a view has one chart grouping and multiple channels. Each channel has a data mapping, a data domain (if the data mapping is non-empty), and a resulting visual output. A view is part of a canvas, and a canvas has a certain position in the semantic space. 

The semantic space has two axes, representing (1) redundancy/compactness (shorthand: compactness axis) and (2) inconsistency/consistency (shorthand: consistency axis). %When the degree of redundancy or inconsistency of a canvas changes, the position along the corresponding semantic axis also changes. 
%These axes allow for reasoning about how redundant and inconsistent a canvas is. 
The compactness axis ranges from redundant to compact, whereas the consistency axis ranges from inconsistent to consistent. For example, if a canvas is made more compact by turning two views into one, the canvas becomes less redundant, thus moving up along the compactness axis. Semantic snapping corresponds to movement along one of these semantic axes.

It is important to note that more consistency and compactness is not always desirable. For example, Qu and Hullman \cite{QuKeepingMultipleViewsConsistent2018} found that in cases, homogenizing axis domains is undesirable due to the extra white space it generates. Conversely, a compact design is not always more readable, or the most ideal for telling a story. Our operations make it possible for the designer to explore this space of alternative designs more rapidly, one semantic axis at a time.
%For example, consider a canvas where two views are using the same color scheme to show different data. Changing the color scheme of one of the views so that they do not conflict would make the canvas more consistent -- hence it would move the canvas position along the consistency axis. 
%Our method lets the user incrementally move a canvas in the semantic space by modifying its overall design, along one axis at a time.
Relations between individual views identify not only where the canvas is currently located, but also which changes (denoted operations) are possible.

\subsection{Algebraic Relations}

% Discuss here what we call a relation, how we identify them 
Relations are explicit specifications of redundancies or inconsistencies between views. Although relations themselves do not indicate whether a design is good or bad, they are available to help the user identify potential problems in their overall visual design. 

Our specification of relations draws both from Qu and Hullman's evaluation constraints~\cite{QuEvaluatingVisSets2016} and a generalization of the principles established in Kindlmann \& Scheidegger's algebraic model of visual design~\cite{kindlmann2014algebraic}. Originally developed in the context of only a single visualization, their model describes the relationships between three elements of the visualization process: the data, the representation of the data, and the resulting visualization. We adopt two principles from this model, which are easily framed within the two high-level constraints stated by Qu and Hullman~\cite{QuEvaluatingVisSets2016}:

\begin{description}[style=unboxed,leftmargin=0cm, noitemsep, topsep=2pt]
    \item [C1] Encode the same data in the same way. A violation of this constraint corresponds to representation invariance in Kindlmann \& Scheidegger's algebraic model, which states: if the data of two visualizations are the same, the resulting visualizations should also be the same. A violation of this is called a \textit{hallucinator} (Table~\ref{tab:semsnap_relations}e).
    \item [C2] Encode different data in different ways. A violation of this constraint has a corollary again in Kindlmann \& Scheidegger's model as an unambiguous data depiction, which states: if the resulting visualizations are the same, the data should also be the same. If this principle is violated, we say that there is a \textit{confuser} (Table~\ref{tab:semsnap_relations}f).
\end{description}

%They describe three principles, two of which we use as a basis for our model: 
%\begin{description}[style=unboxed, leftmargin=0cm, noitemsep, topsep=2pt]
%    \item [(1) Representation invariance] If the data of two visualizations are the same, the resulting visualizations should also be the same. A violation of this is called a \textit{hallucinator}.
%    \item [(2) Unambiguous data depiction] If the resulting visualizations are the same, the data should also be the same. If this principle is violated, we say there is a \textit{confuser}. 
%\end{description}
% YK: Done ish, should hopefully be more clear now, the algebraic "specification" w/o the commutative diagram    

We represent aspects of a single view with the following four elements: the chart grouping (G), a channel (C), the data shown by the channel (D), and the resulting visual output (V).
Typically, the term ``channel" can denote the entire mapping from data to visual output. However, in our case C simply represents the name of the channel, e.g., fill color, and we use the other lower-level elements to concisely specify relations between views as predicate logic expressions as specified in Table~\ref{tab:semsnap_relations}. 
%However, for being able to reason about mappings and retinal outputs more precisely, this more low-level decomposition is necessary. 
%These elements represent information about a single view, and allow for expressing concise rules for identifying relations between different views. 

A single view has a grouping (G), which is the first element in our model. Each view has multiple channels, with (C) referring to a \textit{single} channel, and correspondingly each single channel has a data mapping (D) and a resulting visual output (V). Thus, each view has one (G, C, D, V) tuple \textit{per channel}, where G is always the same, while (C, D, V) is unique to each channel as shown by Figure~\ref{fig:semsnap:relations-conceptual}. 
Consider a bar chart grouped by category, showing average rank on the y-axis. Since $G$=category, $C$=bar height, $D$=average rank, and $V$=y-position, the tuple for the channel is then (category, bar height, average price, y-position) as seen in Figure~\ref{fig:semsnap:relations-conceptual}. If a channel does not have a data mapping, this is expressed as $D=0$. 
%Consider a bar chart grouped by number of cylinders, showing average price on the y-axis. Since $G$=num-cylinders, $C$=bar height, $D$=average price, and $V$=y-position, the tuple for the channel is then (num-cylinders, bar height, average price, y-position). If a channel does not have a data mapping, this is expressed as $D=0$. 

By considering a single view to be a set of (G, C, D, V) tuples (see Figure~\ref{fig:semsnap:relations-conceptual}), we can establish relations between two views by using predicate logic on the tuples and their equalities. When comparing the tuples of two views, we use the same lower case letter to denote equality or inequality. For instance, if two views have the same chart grouping, the relation between $G_1$ and $G_2$ is the identity: $g=1$. Conversely, if the groupings are different, then $g \neq 1$. If a relation exists between the views $A$ and $B$, and between $A$ and $C$, it also exists between $B$ and $C$. 
%, as illustrated in Figure~\ref{fig:semsnap:commutative}.  

A relation exists between two views if there are two tuples (one from each view) that satisfy the predicate logic formula. For example, consider the predicate logic expression of the \textit{multiples} relation: $g=1 \land \exists{c}((c=1 \land \exists{D} \neq 0) \rightarrow d \neq 1)$. This relation exists between two views if there is a pair of channels (one from each view) that satisfy this expression. As illustrated in Figure~\ref{fig:semsnap:relations-conceptual}, the two highlighted views have the same grouping (category), but are showing different quantities on the y-axis (rank vs. score), making the \textit{multiples} expression come true.

As discussed by Qu and Hullman \cite{QuEvaluatingVisSets2016}, two encodings are showing the \textit{same field} when the fields are \textit{semantically} the same. We use this definition. Thus, if two fields are semantically the same, $d = 1$. To confirm semantic sameness, the user is asked to confirm if fields are the same, as seen in Figure~\ref{semsnap:fig:ui_examples}.1b if this cannot be directly determined. 
% Too much detail or ok? Makes style inconsistencies for unmapped channels "ok" within framework
%To this notion of ``same field", we also specify that $d \neq 1$ if both data mappings are empty, but the groupings are different. For example, if the shapes of two views are both colored red, but the groupings are different, this is a potential confuser, since $d \neq 1$. 
We also specify that $d \neq 1$ if both data mappings are empty, but the grouping is different. For example, suppose two pie charts are respectively grouped by gender, and age group, and are both colored red. A sector of the pie chart can then represent either an age group, or a gender, yet they are colored the same. This is a potential confuser since each are showing different data, but are colored the same.

We define that the stroke color channel of charts without filled shapes (e.g., a line chart), and the fill color channel for a any chart with a fill (e.g., bar chart, scatterplot), is the same. For example, in the example shown in Figure~\ref{fig:teaser}.1 we see a line chart and a scatter plot both using the color red. With our notion of channel equality, $c = 1$ since the stroke color of the line chart is the same as the fill color of the scatterplot. Furthermore, they are grouped differently ($g \neq 1$), and both of the color channels are not mapped to data. As a result of these two factors, the data mappings of the two channels are seen as different: $d \neq 1$.

The degree of redundancy and compactness in a view can be measured by the number, and severity, of detected relations. A design is more compact if it has fewer relations indicating redundancy, and more consistent if it has fewer relations indicating inconsistency. For our method, it is only necessary to know that a relation exists, and that it can be resolved. However, generating a quantitative score from these relations would be possible, and useful for many other problems. These relations are specified and visually summarized in Table \ref{tab:semsnap_relations}. We discuss each of these relations in detail in the remainder of this section.

\begin{figure}[tb]%[t]
    \centering
    \includegraphics[width=\linewidth]{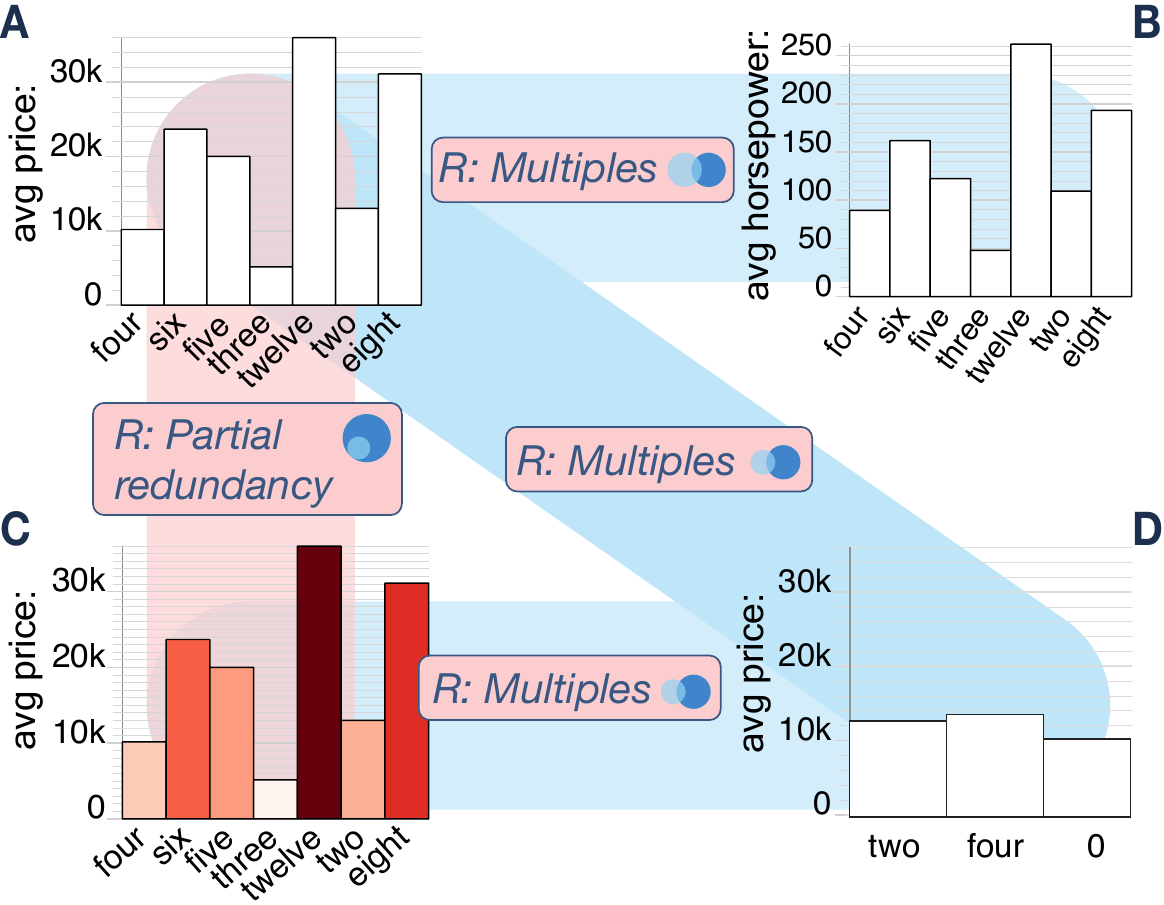}
    \caption{These four figures illustrate all degrees of redundancy. There is a multiples relation between (a) and (b) since they both have the same grouping, but are showing different data via the bar height channel. Between a and c we see there is a partial redundancy, since (a) is showing the exact same data as (c), but (c) is also showing more data via the fill color. (a) and (d), as well as (c) and (d) are differently grouped multiples showing the same data via the bar height.}
    \label{semsnap:fig:redundancies}
\end{figure}

\begin{description}[leftmargin=0cm,style=unboxed]
\item[R1: Full Redundancy.] %
If two views are showing exactly the same data, there is a full redundancy relation between them. The full redundancy relation is present when two views have the same grouping ($g=1$) for all channel pairs ($\forall{c}$). If the channels are the same ($c=1$), then they also show the same data ($d=1$). For example, if two bar charts are both grouped by number of cylinders ($g=1$), and their bar height is mapped to average price, then ($c = 1 \rightarrow d = 1$) is true, i.e., there is a full redundancy relation between them.

\item[R2: Partial Redundancy.] %
%Two views $D1$ and $D2$ are partially redundant if $D2$ is showing all data shown by $D1$, and there is some data not shown by $D1$. 
%
Two views $A$ and $B$ are partially redundant if $A$ is showing all data shown by $B$, as well as some data not shown by $B$. More formally, two views are considered partially redundant if they have the same grouping ($g=1$), and for all pairs of channels showing different data ($c=1 \land d \neq 1$), one of the channels is unmapped, and all the unmapped channels consistently belong to the same view ($\exists!{D}=0$). 
For example, consider two bar charts, both grouped by number of cylinders ($g=1$) and with bar height mapped to average price, but with one chart also indicating the number of cylinders via its fill color channel. When comparing the fill color channels of the charts ($c=1$), we see that they have different data mappings ($d \neq 1$), and that one of them is not mapped to anything ($\exists!{D} = 0$). Thus, all the data shown by the one chart is also shown by the other chart.

\item[R3: Multiples.] %
There are two kinds of multiples: (1) views with same grouping but different data, or, conversely, (2) views with different groupings but same data. As an example of the former, suppose two equally grouped bar charts showing a different quantity via the bar height channel as illustrated in Figure \ref{semsnap:fig:redundancies}a and b. Multiples with different groupings could for example be two differently grouped bar charts showing the same aggregated dimension via the bar height channel (see Figure \ref{semsnap:fig:redundancies}a and d). The multiples relation is specified more precisely in Table \ref{tab:semsnap_relations}c-d.
%Multiples may exist both with, or without the same grouping. If they have the same grouping
%Multiples, both with and without the same grouping, are identified by requiring that each view has the same channel ($c=1$) mapped to different data($d\neq 1$), and that both channels are in fact mapped to data ($D_1 \neq 0$) . 
%When two views are both showing data not shown by the other view, they might be \textit{multiples}. Our use of the term: multiples is based on the established use of small multiples. Small multiples is an arrangement of similar views, using the same scale and axis. The multiples relation identifies whether two views are similar in a small-multiples fashion. Multiples are similar, yet unique. They show some of the same data, yet each view is also showing something not shown by other views. There are two kinds of multiples: (1) same grouping but different data, or conversely, (2) different grouping but same data. As an example of the former, suppose two similarly grouped bar charts showing a different quantity via the bar height channel as shown in Figure \ref{semsnap:fig:redundancies} a and b. Multiples with different groupings could for example be two differently grouped bar charts showing the same aggregated dimension via the bar height channel (See Figure \ref{semsnap:fig:redundancies} a and d). 

% YK: C1 violation means same data shown differently, not necessarily a
\item[R4: Hallucinator.] %
Corresponding to Kindlmann and Scheidegger's model, a hallucinator is present when the same data are shown in different ways. A hallucinator exists on a canvas if two views with the same chart grouping ($g=1$), have a common channel ($c=1$) showing the same data ($d=1$), but with different visual output ($v \neq 1$). %However, if $d \neq 1$, but the data shown is semantically the same, there is also a hallucinator. In other words, if $d=1$ and all the other conditions hold, there is a hallucinator for certain. Conversely, if $d \neq 1$ there is a \textit{potential} hallucinator.

\item[R5: Confuser.] %
%When the visual outputs are the same, but the data is different, there is a confuser.%When the data are different, but the visual outputs are consistent, there is a data inconsistency. %If two views are showing different data via the same channel, with the same visual output, there is a C2 violation. 
A confuser exists on a canvas if the same channel ($c=1$) of two views has the same visual output ($v=1$), but different data mappings ($d \neq 1$). As an example, consider charts using the same fill color (for example, reds) to show different data. 
\end{description}

Relations identify redundancies, inconsistencies, and alternative design opportunities. They are detected by iterating over all view permutations and checking whether the permutation satisfies the predicate logic expression that corresponds to the rule. If the expression is satisfied, the relation exists between the views. Each relation has corresponding "resolutions" -- operations which resolve the given relation by altering one or more of the affected views. 

%Suppose we have two views that are multiples, showing different data dimensions ($d \neq 1$) through the same channel ($c=1$) (See Figure \ref{semsnap:fig:integrate_examples}a versus \ref{semsnap:fig:integrate_examples}b). If the two data dimensions are showing the same semantic quantity, for example the amount of sales, but the data domains are different, there is a confuser.
%-------------------------------

\subsection{User Operations}
Operations resolve potential problems between views of a canvas. The main idea behind each operation is to resolve a certain relation by changing or removing one or more views. At a low level, operations can, for example, transfer a data mapping from one view onto another, or replace two views with another view showing the same data. Our set of operations do not exhaustively express all possible combinations of low-level changes, but serves to demonstrate the wide range of possible operations to navigate the semantic space of revised designs.

Our model specifies the following classes of operations, which we summarize in Table \ref{tab:semsnap_relations} along with associated their associated relation(s).
%\begin{description}[style=unboxed,leftmargin=0cm, noitemsep, topsep=2pt]
%    \item [O1] Delete: Remove a view in case of full redundancy between views
%    \item [O2] Homogenize: Make two views more similar in case of multiples or hallucinator
%    \item [O3] Differentiate: Make two views more different in case of confuser
%    \item [O4] Integrate: Integrate views in case of partial redundancy or similarly grouped multiples
%\end{description}

\begin{description}[leftmargin=0cm,style=unboxed]
\item[O1: Delete.] %
The first operation is the most simple. If there is a \textit{full redundancy} relation between two views, the user may delete one view. With one of the views removed, the formula for \textit{full redundancy}, $g = 1 \land \forall{c} (c=1 \rightarrow d = 1)$, will not evaluate to true for that pair of views, since the pair no longer exists.

\item[O2: Homogenize.] %
The \textit{homogenize} operation resolves \textit{hallucinator}, as well as \textit{multiples} relations, where the same data are shown differently. On a high level, the \textit{homogenize} operation makes dissimilar views more similar. For example, consider two bar charts that are showing the same data dimension with a different color scheme (hallucinator), as in Figure~\ref{semsnap:fig:example_3}.1a, or that are using different data domains (multiples), as shown in Figure~\ref{semsnap:fig:example_2}.2. The \textit{homogenize} operation resolves these conflicts by making the visual outputs, or data domains equal for the two views. If visual outputs are made equal, the $v \neq 1$ portion of the \textit{hallucinator} will evaluate to false and thus remove the relation. For multiples with different domains, equalizing the domains will make the views consistent. If only the data domains are different, the operation is presented to the user as \textit{homogenize} data. When the visual outputs differ, the user will see the operation as \textit{homogenize} style, although it also implicitly homogenizes the data domains.
%We identify \textit{hallucinators} by looking for channels in conflict, i.e., channels of the same type ($c=1$) showing the same data ($d=1$), but with different visual outputs ($v \neq 1)$. The \textit{homogenize} operation resolves these conflicts by making the visual outputs equal. If the outputs are made equal, the $v \neq 1$ portion of the \textit{hallucinator} will evaluate to false. This removes the relation from the design. The visual output is also affected by the data domain. Thus, the \textit{homogenize} operation is applicable to both of these components. If only the data domains are different, the operation is presented to the user as \textit{homogenize} data, otherwise. When the visual outputs differ, the user will see the operation as \textit{homogenize} style, although it also implicitly homogenizes the data domains.
 
 %\textbf{Homogenize Visual Outputs}
 %If the same data is shown through the same channel, but with different visual outputs (Table \ref{tab:semsnap_relations}e), this can be resolved by making the visual outputs the same. For example, consider multiple charts showing the same data through fill color with different color schemes, the solution to this would be to make the color scheme uniform across all charts.
 
 %\textbf{Homogenize Data Domains. }When the same data is shown with the same channel across multiple charts, but with different data domains (Table \ref{tab:semsnap_relations}f), this can be resolved by homogenizing the data domains. 
 
 %For example, ...TODO get a figure too maybe?

\item[O3: Differentiate.] %
The \textit{differentiate} operation addresses a \textit{confuser}, where two views show different data in the same way. This is achieved by making the views to show different data in different ways~(Table~\ref{tab:semsnap_relations}f). For instance, if different data are shown using the same color scheme, the \textit{differentiate} operation will assign different color schemes to the views. An example of this operation could take two views showing different data, such as age and income, where both views are mapped to the color red. This is ambiguous. Our solution is to use a different color scheme for one of the views. When the visual outputs are made different, i.e., from $v = 1$ to $v \neq 1$, the formula for a \textit{confuser}, $c=1 \land d \neq 1 \land v = 1$, will evaluate to false, since $v \neq 1$.

\item[O4: Integrate. ] %
The \textit{integrate} operation resolves redundancy to create a visually compact canvas, and can be used to resolve \textit{partial redundancies} and certain \textit{multiples} relations. With a \textit{partial redundancy} relation, there are two possible solutions: delete the view showing the least data, or \textit{integrate} the ``missing'' mapping into this view while deleting the other~(Table \ref{tab:semsnap_relations}b). Views sharing a \textit{multiples} relation where the data grouping is the same~(Table \ref{tab:semsnap_relations}c) can be \textit{integrated} in several ways. %In each of these integration types, the act of combining two charts removes the spatial redundancy between the charts and thereby their \textit{multiples} relation.
Since the multiples relation can only exist between two views, the act of combining these views by integration also removes the relation from the canvas.
Views that are highly semantically similar are sensible to \textit{integrate}, provided that: (1) the chart type is the same, and (2) if $d=1$ for the channel representing the x-axis. There are four ways to perform this \textit{integration}: \textit{overlay, group, stack,} and \textit{mirror}. 
%Both chart type and semantic or strict equality between channels representing axes determine which operations are available. All integrations of multiples require semantically equal y-axis channels (for example, bar height) and strictly equal x-axis channels (for example, bar order), or the opposite.
%
\textit{Overlay} integrates multiple views into the same coordinate system. This operation can be applied to scatter plots and line charts. Figure~\ref{fig:teaser}.2* shows an example of an \textit{overlay} operation when applied to a line chart. The \textit{mirror} operation can be applied to line charts, area charts, and bar charts. This operation first aligns the two views and then mirrors one of them, causing their marks diverge from a common origin in a manner similar to violin plots. We demonstrate an example of this in Figure~\ref{semsnap:fig:ui_examples}.2b. Similar to the group operation, the \textit{stack} operation stacks views into a single view, turning, for instance, a set of bar charts into a stacked bar chart as shown again in Figure~\ref{semsnap:fig:ui_examples}.2c. The \textit{group} integration bundles several views into one single view. It can, for example, turn multiple bar charts into a grouped bar chart, as illustrated in Figure~\ref{semsnap:fig:ui_examples}.2d. 

\end{description}

%If both axes are semantically equal, but not strictly equal, the overlay operation is still available for scatter plots (for an example, see Figure TODO).
Operations make high-level changes to the canvas, making it more compact or more consistent. For an operation to be applicable to a design, a certain relation must exist. When the user selects a view in the interface, our method reveals available operations to resolve a given relation. When the operation is performed, the corresponding relation is addressed.

%-----------------------------------------------
\subsection{Snapping Algorithm}

\begin{lstlisting}[float=tb,language=Python, numbers=right, basicstyle=\small, belowskip=-0.5 \baselineskip, caption={Pseudocode of the general execution flow of semantic snapping.} \label{lst:semsnap_controlflow}]
def findRelations(views):
    byView = { view: [] for view in views }
    subsets = permutations(views) 
    for view1, view2 in subsets:
        for relationFn in allRelations:
            if relationFn(view1, view2):
                for view in subset:
                    byView[view].append({ 
                        'subset': [view1, view2], 
                        'relation': relation })
    return byView
    
def semanticSnap(views):
    relationsByView = findRelations(views)
    view = userInput() # User selects a view
    relations = relationsByView.get(view)
    operations = [ findOperation(r) for r in relations ]
    display(operations) # User sees operations
    selectedOperation = userInput() # User selects
    newViews = selectedOperation.execute()
    return newViews
    
\end{lstlisting}
\begin{lstlisting}[float=tb,language=Python, numbers=right, basicstyle=\small, belowskip=-0.5 \baselineskip, caption={Pseudocode of an example relationFn, invoked at Listing~\ref{lst:semsnap_controlflow} line 6, modelling a hallucinator as specified in Table \ref{tab:semsnap_relations}e.} \label{lst:semsnap_c1violation}]
def isHallucinator(view1, view2):
    if isSameGrouping(view, view2):
        pairs = findChannelPairs(view1, view2,{   
            'd': 1, # same data
            'v': 0, # different visual output
            'mappedToData': 1  }) # D != 0
        return pairs.length > 0 
    return 0
\end{lstlisting}

The goal of our approach is to provide the user with a set of available design-altering operations upon selection of a single view. The outlined algorithm in Listing \ref{lst:semsnap_controlflow} achieves this goal by identifying all relations and mapping them to operations for any selected view. When an operation is selected, a revised set of views is generated. The relations correspond to the descriptions in Table \ref{tab:semsnap_relations}, and can in practice be modeled as constraints or functions. 

The first step of the algorithm is to identify all relations between all subsets of views. The logic of each relation is outlined in Table \ref{tab:semsnap_relations}, and is mapped to a relation function that takes in two views, and returns \lstinline{1} if the relation exists between the views, or \lstinline{0} if the relation does not exist between the views, as exemplified in Listing \ref{lst:semsnap_c1violation}. Consider line 5 in Listing~\ref{lst:semsnap_controlflow}. Here we loop over each \lstinline{relationFn} (relation function), and invoke it using two views as arguments. If this invocation returns \lstinline{1}, the relation exists between the two views. When relations are identified for all views, they are grouped by the views they affect. When the user selects a view, the view's relations are looked up and used to identify which operations are possible. Line 17 of Listing~\ref{lst:semsnap_controlflow} illustrates how relations are mapped to corresponding operations. When an operation is executed, a new set of views is generated and displayed to the user. With this new set of views, the algorithm is re-run, recomputing relations and potential operations.

%-----------------------------------------------

%-----------------------------------------------
\section{Workflow \& Implementation}%\section{User Interface \& Implementation}
Our method improves and refines canvas designs incrementally. In order to create a canvas, single visualizations must also be generated. While the creation of single visualizations is not a part of our method, we used the existing Visception visualization editor environment~\cite{Kristiansen-2020-VIV} as a basis to realize and demonstrate our method.
Our semantic snapping interface enables the user to build a canvas using simple drag \& drop operations from a visualization gallery and to optimize the design with semantic snapping step by step. In order to build a canvas, the user places individual views into a grid layout and is presented with a set of potential operations at every step. While browsing the operations, the user is presented with information about what they do and what potential problems in the design they resolve.  

%\subsection{User Interface}
%\begin{figure}
%    \centering
%    \includegraphics[width=\linewidth]{pictures/pictures_v2/interface.pdf}
%    \caption{The \textit{Semantic Snapping} interface. It consists only of two views, the singles view to the right from where the user can drag visualizations, and place them into the multiples view to the right.}
%    \label{semsnap:fig:ui_overview}
%\end{figure}

We use two primary views in our interface: the \textit{singles view} and the \textit{canvas view}. Both are shown in Figure~\ref{semsnap:fig:ui_examples}.1. 
The \textit{singles view} is a view from where the user can drag single visualizations into the \textit{canvas view}. Each tile in the singles view represents a data source, which, when clicked, expands to more tiles--one for each single visualization of that dataset. We highlight two of these single visualization tiles in Figure~\ref{semsnap:fig:ui_examples}, which have been dragged into the canvas view. 

\subsection{Workflow}%

\begin{figure}[tb]%[!htb]
    \centering
    \includegraphics[width=\linewidth]{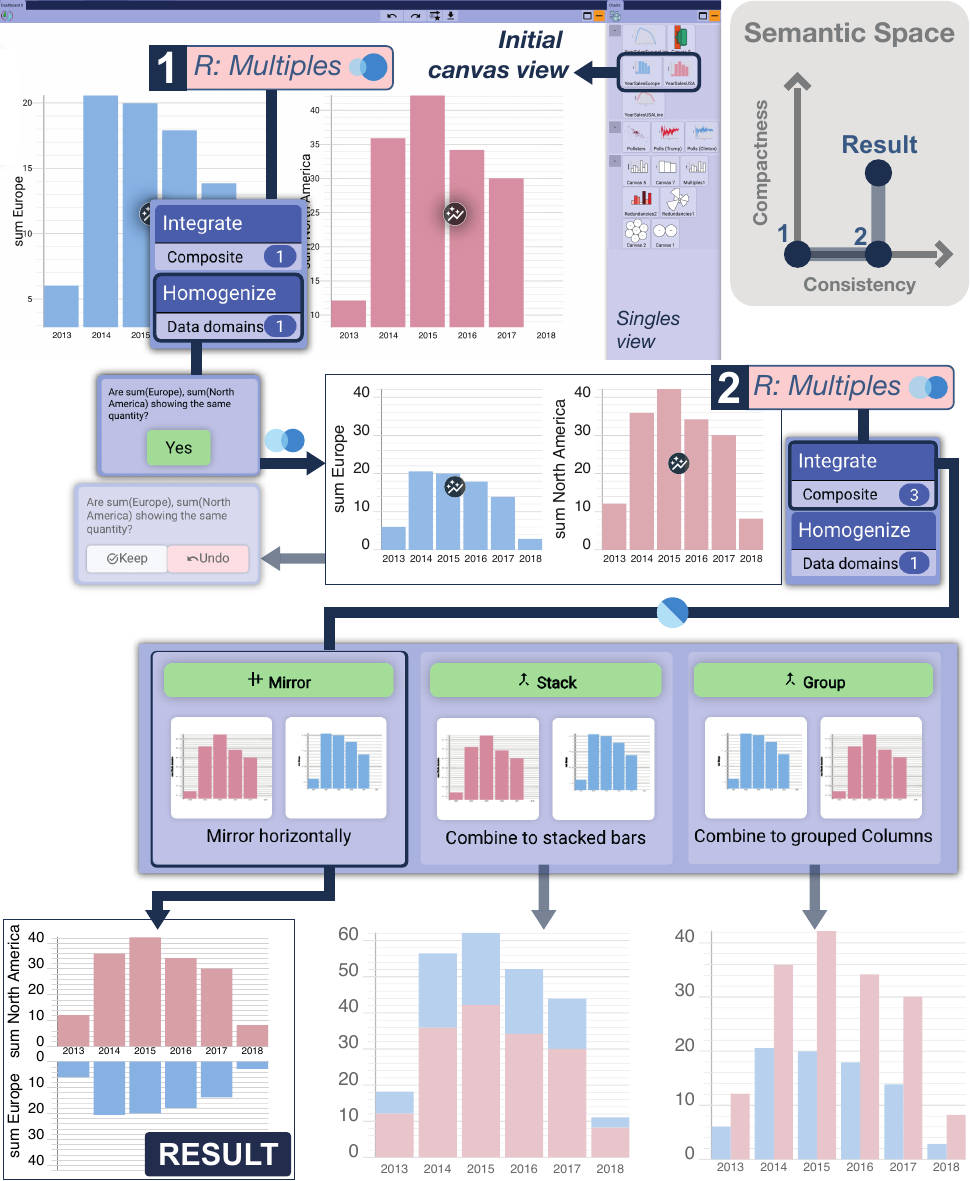}
    \caption{Semantic snapping interface and workflow. The semantic snapping interface presents the user with a canvas to place single views within a larger layout. A clickable button on top of each view exposes the possible operations available to resolve a relation between two views, in this case a \textit{multiples} relation (1). The user  \textit{homogenizes} the two views after confirming that the y-axes are semantically the same. This step represents a move towards increased consistency in semantic space. The user may choose to keep or undo the result of any performed operation. On execution of an operation, we recompute the set of possible operations. The user may next perform any one of three available \textit{integration} operations to resolve a \textit{multiples} relation in this view (2). The user selects the \textit{mirror} operation to increase the compactness of the visualization in semantic space. 
    }
    \label{semsnap:fig:ui_examples}
\end{figure}

The main workflow of using semantic snapping is integrated into the \textit{canvas view}. 
We demonstrate this workflow in Figure \ref{semsnap:fig:ui_examples}. As the user is constructing a canvas with multiple views, our approach detects relations and makes the corresponding operations available along every step of the way. Whenever an existing view is added to the design, or modified by an operation, relations are re-detected and the corresponding operations are updated. In order to see possible operations, the user clicks on the view of interest. When the view is clicked, a menu appears, showing how many operations are available per category (homogenize data, homogenize style, differentiate, and integrate). Only categories with available operations are displayed. The user can then click on a category and see all available operations as tiles. Each operation tile informs the user of the potential problem and its solution. %In cases where the operations validity depends on the semantics of some underlying data. For example, it makes sense to homogenize the data domains if the underlying data shown is semantically the same. To accommodate this, some operations will ask the user if some data dimensions are semantically the same. 
Consider the canvas in our workflow example where there is a \textit{multiples} relation between two bar charts (Figure~\ref{semsnap:fig:ui_examples}.1), showing sum of sales in Europe, and North America. To verify that the fields are semantically equal, the operation tile will ask the user "Are sum(Europe) and sum(North America) representing the same quantity?". If the user clicks "Yes", the domains are made consistent using a \textit{homogenize operation}. When any operation is executed, the user is given the option to undo or keep it as shown in~Figure~\ref{semsnap:fig:ui_examples}.2. If "keep" is clicked, the current canvas is re-evaluated and the user can proceed to explore other operations, add new views, or otherwise customize the setup.

%As an example of the user being asked for information, consider a design where there is a potential data domain inconsistency, the operation will for example ask the user "Are sum(Europe) and sum(North America) representing the same quantity?" (See Figure \ref{semsnap:fig:ui_examples}c). If the user clicks "Yes", the domains are made consistent. When any operation is executed, the user is given the option to undo it, or keep it as shown in Figure \ref{semsnap:fig:ui_examples}e. By default it is kept. If "keep" is clicked, all new solutions are recomputed.

\subsection{Implementation}
We implemented semantic snapping within the framework of Visception~\cite{Kristiansen-2020-VIV}, an application written in Javascript ES6 using VueJS for user interface components and D3 for SVG rendering. The underlying framework of the authoring tool was leveraged to realize the relations and operations to support semantic snapping.

The relations are specified as functions taking in two views as parameters, returning true if they match the given relations. Operations are also defined by functions that take in a set of views, and a detected relation. From this set of views and the detected relation, we can infer what operations are possible, and also take into account the specific chart type and other edge cases. 
When the user modifies a design, a pipeline of four steps is run. First, all relations are detected for all sets of relations as shown in Listing \ref{lst:semsnap_controlflow}, and the relations are stored so that they can be looked up on a per-view basis. When the detection is done, the editor is ready for the user to specify which view to change. When the user clicks on a view, all relations and corresponding sets of views are looked up, and all possible operations are computed. When the user selects an operation, it is executed, and the existing set of views is modified, and relations are recomputed.

% How rules are mapped to javascript
% How control flow is
% Drag&drop
% ... 

% What it is written in (ES6)
% Screenshot of interface
% Screenshot of operations
% Screenshot of operation description
% Underlying thingy behind UI etc
% What it is built on (already working editor)
% How the rules mapped nicely to rules in Javascript
% How the interface is built up 
% (dragging charts to place them, clicking on one chart, seeing operations, ..)

%-----------------------------------------------
\section{Case Studies}

We next demonstrate our semantic snapping method workflow in three case studies. These studies include data from the 2016 US Election Results, Nightingale's historic Soldier Morbidity \& Mortality, and a COVID-19 dataset. We selected these particular datasets as they are both representative of the type of data we expect to be used for our approach, as well as for their familiarity and applicability to the visualization community. Each case study represents a possible pathway through semantic space from an initial to a more compact and consistent design. 
We illustrate such pathways through semantic space with a semantic space map positioned in the upper right of each associated figure. 
%We illustrate this path through semantic space, alongside other possible paths, in a semantic space map that we position in the upper right of each associated study figure. 
%We selected the set of examples in order to cover most operations, and present them in an understandable fashion. 

\subsection{2016 Election Results}
In this case study we demonstrate a user flow that identifies \textit{confuser} and \textit{multiples} relations that are resolved via \textit{differentiate} and \textit{integrate} operations. We also use this study case to demonstrate a flexible workflow whereby the user may perform and then revert an operation to arrive at their preferred final design.

In Figure~\ref{fig:teaser} we see an initial canvas comprised of three views depicting data from the 2016 US Election. These views show election polls over time for the two main candidates (left, top view: Democrats, bottom view: Republicans), as well as average pollster ratings for the two candidates (right scatter plot view). We localize our position in semantic space at the origin (pos.~1) in the map in the upper right of Figure~\ref{fig:teaser}. We quickly identify a \textit{confuser} relation between the bottom line chart and the right scatter plot~(Figure~\ref{fig:teaser}.1). This is because the color channels of the two views are using the color red as visual output. This is particularly misleading in the right scatter plot view, where each dot represents a pollster, since red may indicate that all pollsters are advocates for the Republican party. Since these charts are using the same visual output to represent different data domains, our model recommends a \textit{differentiate} operation to change their respective visual outputs. We change the color of the scatter plot to green, as this is color is more neutral. We keep the red color in the lower left view; this makes sense to remain red, as this is the color of the US Republican Party. In our semantic map we have increased the consistency of our canvas and are now at pos.~2. 

We next observe a correspondence between the two leftmost views. These views share the same x-axis, but show a different quantity on the y-axis (bottom view: avg. Trump, top view: avg. Clinton). In other words, they share a \textit{multiples} relation. We consequently can \textit{integrate} them to produce a more compact visualization using the \textit{mirroring}~(Figure~\ref{fig:teaser}.2) or \textit{overlaying operation}~(Figure~\ref{fig:teaser}.2*). Integrating these charts additionally produces a more consistent visualization, because both \textit{overlay} and \textit{mirror} perform an implicit axis homogenization step.
To mirror or overlay, we simply select and execute either operation. In this case we first try \textit{overlay}~(Figure~\ref{fig:teaser}.2*). However, while the result is very compact, it is difficult to read. We choose to do a different \textit{integrate} operation to resolve the \textit{multiples} relation. We revisit the available operations for this relation and select this time to \textit{mirror} the two views~(Figure~\ref{fig:teaser}.2). This path in semantic space leads us to an equally consistent, while slightly less compact visualization. The resulting chart composition, however, is easier to read, which illustrates the flexibility of our approach in incorporating user goals and decision processes.

\subsection{Nightingale Soldier Morbidity \& Mortality in 1858}
In this case study we use the popular Nightingale solider morbidity \& mortality dataset to illustrate the use of additional operations to resolve \textit{multiples} between canvas views.
\begin{figure}[tb]
    \centering
    \includegraphics[width=\linewidth]{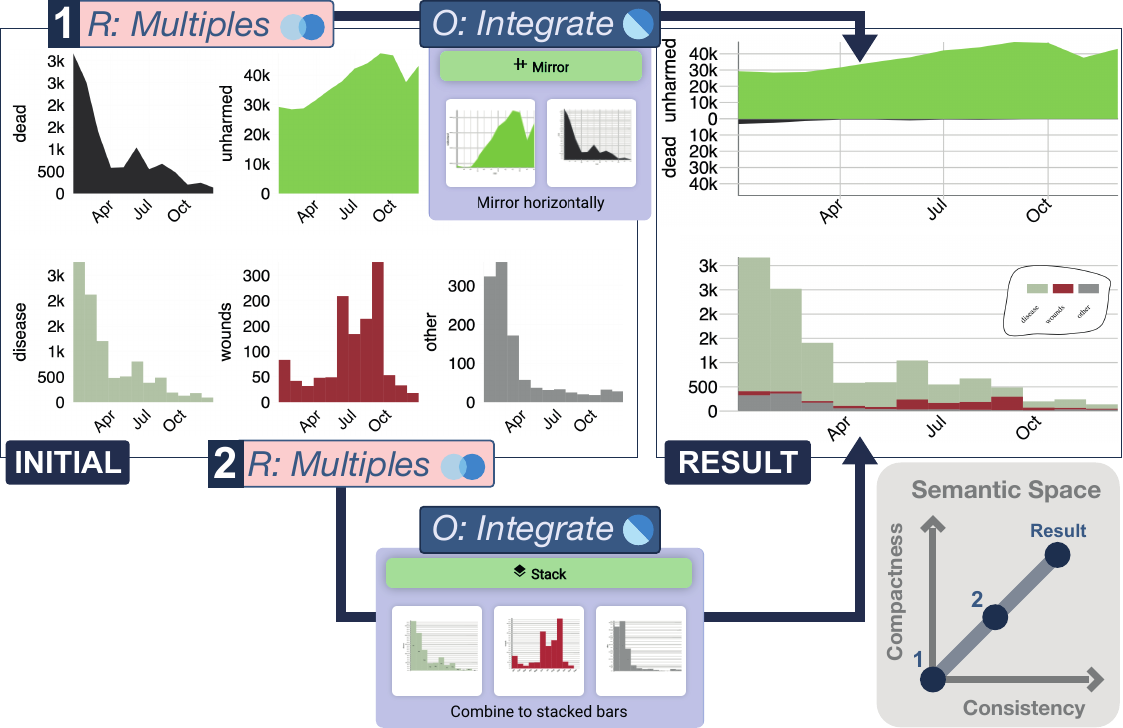}
    \caption{Case study workflow demonstrating semantic snapping to resolve two \textit{multiples} relations in the canvas depicting 1858 solider morbidity \& mortality from the Nightingale dataset. This end result is a semantically consistent and compact visualization.}
    \label{semsnap:fig:example_2}
\end{figure}

Figure~\ref{semsnap:fig:example_2} shows the fate of British soldiers in the year of 1858 in the Crimean War. In our traversal through semantic space we begin again at the origin in our semantic map at the upper right (pos.~1). The top two views of the initial canvas display area charts. The left view plots the number of soldier deaths over time while the right plots the number of unharmed soldiers over time. Because these two views comprise the same grouping (soldier morbidity \& mortality) with a different data domain plotted onto the y-axis, we can say that these views share a \textit{multiples} relation~(Figure~\ref{semsnap:fig:example_2}.1). We can compact the views by \textit{mirroring} (integrating) the views. When the charts are mirrored, the domains are also implicitly homogenized, which additionally increases the consistency of the resulting chart. This brings us to pos.~2 in our semantic map. 
It would also be possible to keep the multiples relation, and only homogenize the data domains on the y-axis.
%While the integration makes the domains the same, it is also possible to keep the charts as multiples and only make the domains the same. This is possible because the views share a \textit{hallucinator} relation, which is there since y-axes are using different data domains, and thus have different visual outputs. 
However, our choice to compact the views with mirroring illustrates that, while our model can show potential problems in a canvas, it is ultimately up to the user to decide how they wish to design their visualization.
%This action makes a \textit{hallucinator} relation clear. This arises because our \textit{mirroring} operation has combined different data domains to share the same visual output (plotted onto the y-axis) of this view. This potential \textit{hallucinator} would have been possible to avoid: we could have instead kept the two views as \textit{multiples}, and simply made the views more consistent by homogenizing the data domains.
%This choice would have taken us down the light blue path in our semantic map. However, our choice to compact the views illustrates that, while our model can show potential problems in a canvas, it is ultimately up to the user to decide how they wish to design their visualization.

We may also compact the three bar chart views arrayed along the bottom of the canvas that show different causes of soldier death~(Figure~\ref{semsnap:fig:example_2}, bottom of initial canvas). Each of the three views plots by number of deaths caused by disease, wounds, or other, respectively, over time. Because these views share the same grouping (cause of solider death) and data domain on the x-axis (time), but their data domain plotted to the y-axis is different, we identify a \textit{multiples} relation~(Figure~\ref{semsnap:fig:example_2}.2). However, by looking at the data, we also know that the y-axes represent the same \textit{semantic} quantity -- i.e., number of deaths. As a consequence of this, they can be integrated via a \textit{group} or \textit{stack} operation. Since our goal is to produce a maximally compact visualization, we choose to \textit{stack} the views. This step additionally homogenizes the data domains for increased consisstency. This brings us to the result point in our semantic map. 

\subsection{COVID-19 in Germany}

\begin{figure*}[tb]
    \centering
    \includegraphics[width=1\linewidth]{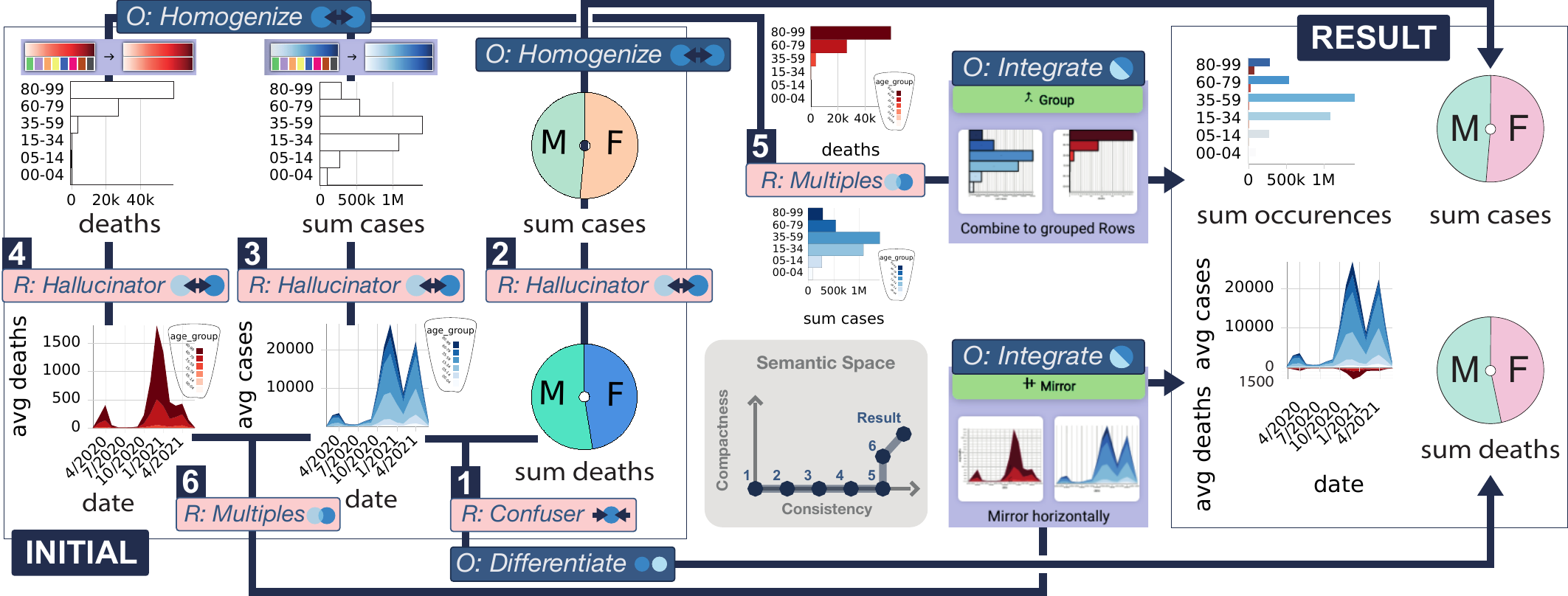}
    \caption{Case study workflow demonstrating semantic snapping to resolve a COVID-19 dashboard with a \textit{confuser}, several \textit{hallucinators} and two \textit{multiples} relations. We resolve these relations via a series of operations that include \textit{homogenize}, \textit{differentiate}, and \textit{integrate}.
    }
    \label{semsnap:fig:example_3}
\end{figure*}

COVID-19 dashboards are now ubiquitous in society with great importance for public health. However, integration of numerous charts to demonstrate various data aspects in a dashboard may introduce numerous possible conceptual and perceptual pitfalls. For our final case study we demonstrate the ability of our approach to assist in resolving the complexities of creating a semantically consistent COVID-19 dashboard. We demonstrate an overview of this workflow in Figure~\ref{semsnap:fig:example_3}.

The initial layout as shown in the central part of Figure~\ref{semsnap:fig:example_3} shows six charts. The first chart column shows COVID-19 deaths grouped by age, where the top bar chart represents total deaths and the bottom streamgraph indicates deaths over time. The second column displays COVID-19 cases that are again grouped by age, with the top bar chart indicating summed cases while the below streamgraph shows case load over time. The rightmost column shows two pie charts grouped by gender, where the top chart shows cases while the bottom shows deaths. Our goal is create a dashboard using multiple chart types that clearly presents the COVID-19 cases and deaths distributed by age group and gender in Germany. 

As in the prior case examples we have a number of different routes through which we can traverse the semantic space, as indicated in map in the lower middle of Figure~\ref{semsnap:fig:example_3}. In this case study we describe the navy blue indicated route, beginning with the \textit{confuser} that our system identifies between the pie chart showing COVID-19 deaths and the COVID-19 cases over time chart (Figure~\ref{semsnap:fig:example_3}.1). This is a \textit{confuser} because the female segment in the pie chart uses the same blue as for the color mapping in the chart showing cases grouped by age over time. We perform the suggested \textit{differentiate} operation to clarify the different groupings by changing the color mapping of genders to a light green for males and pink for females. This increases consistency in semantic space. We next resolve the \textit{hallucinator} relation between the two pie charts by \textit{homogenizing} the color mapping of both charts so the cases chart receives the same green and pink color mapping to males and females, respectively, for increased consistency in semantic space (Figure~\ref{semsnap:fig:example_3}.2). 

Two additional \textit{hallucinators} exist, one between the age-grouped COVID-19 cases charts (Figure~\ref{semsnap:fig:example_3}.3) and the second between the age-grouped COVID-19 deaths charts (Figure~\ref{semsnap:fig:example_3}.4). Each are classified as \textit{hallucinators} because the chart data and groupings are identical but they do not share the same color mapping. We resolve the first \textit{hallucinator} between the two case charts with a \textit{homogenize} operation that applies the same continuous blue color mapping in the streamgraph to the bar chart. We resolve the second \textit{hallucinator} the same way for the deaths chart, by applying the continuous red color mapping in the deaths over time streamgraph to the corresponding bar chart. Each of these operations sequentially improves consistency in semantic space. 

We may compact our dashboard visualization by resolving two \textit{multiples} relations that our system identifies. The first \textit{multiples} relation exists between the two bar charts, which we \textit{integrate} into a single grouped row chart (Figure~\ref{semsnap:fig:example_3}.5). A second compacting step in semantic space \textit{integrates} the two streamgraphs in a mirroring operation to resolve their \textit{multiples} relation (Figure~\ref{semsnap:fig:example_3}.6). In both \textit{integration} steps the system implicitly \textit{homogenizes} the data domains as well. The resulting COVID-19 dashboard in the right of Figure~\ref{semsnap:fig:example_3} is a much more compact and semantically consistent visualization with the aid of our approach.

%-----------------------------------------------
\section{Discussion \& Limitations}

We realized our method by implementing and embedding it into the Visception visual authoring system~\cite{Kristiansen-2020-VIV}. For specifying rules, it is necessary to be able to retrieve detailed information about each channel mapping, as well as the chart type and grouping. We believe this information should be accessible in most frameworks. Specifying chart operations requires more knowledge about the underlying architecture and programming interface. For instance, our framework had support for nesting visualizations, which was highly useful for generating grouped and stacked charts. For example, two bar charts grouped the same, showing two different quantitative attributes can be grouped into a stacked bar chart, where the outer bar is grouped the same, and the inner grouping is one bar for each of the two attributes. In principle, however, we believe that our approach is applicable to a variety of different visualization systems, even if the specifics of how individual operations are implemented will differ.
At present, our prototype only supports a limited number of common chart types: line charts, bar charts, pie charts, scatter plots, and streamgraphs. For addressing a wider range of charts, we believe that a framework for unifying the reasoning about these charts could allow for a more generally applicable realization of semantic snapping. We believe that frameworks that allow for expressing and modifying charts on a general level are ideal for implementing our semantic snapping concept.

At present, a canvas is limited to views of a single tabular dataset. For dealing with more advanced multi-table setups, our approach would have to be built on top of an additional abstraction over these different data topologies. Such an abstraction layer would enable a more general implementation of semantic snapping that could also facilitate the incorporation of other dataset types such as network data. Likewise, techniques such as interactive linking \& brushing and crossfiltering are frequently used in multi-view visualizations but currently not explicitly supported in our framework, which also represents an interesting challenge for future research.
 %Some edge cases were also handled. For example, the stroke color of a line chart can be seen as the fill color of another chart in terms of conflict as seen in Figure~\ref{fig:teaser}.1a. 

The layout of the views is an important factor in an overall design, which is currently not addressed in our approach but is definitely worthy of further investigation. It would be possible to specify more advanced relations by incorporating the spatial arrangement of individual views. For instance, if two views are sufficiently spatially separated, a \textit{confuser} could be classified as less severe. Likewise, taking into account spatial arrangement could extend the space of operations as, for example, a \textit{differentiate} operation could move views further apart or even add graphical separators or visual groupings. This is an important direction for future research. Related to this, since currently the number of possible operations is sufficiently small, we do not perform any explicit sorting. However, a larger number of possibilities would necessitate to incorporate an appropriate mechanism for prioritizing operations. We believe that such a sorting of potential revisions using for example Qu \& Hullman's effectiveness preservation score \cite{QuEvaluatingVisSets2016} would be useful when there are many potential solutions.

While the operations of our method alter the design and resolve potential inconsistencies, it would provide more flexibility and design freedom if they were customizable. For example, the mirror operation could be parameterized by letting the user decide the spacing between the views and the placement of the labels. A general assumption of our method is that the existing views are already well-designed individually. However, when a view is placed into a multi-view design, the aspect ratio and size will change. Keeping font sizes and other styles consistent across a design becomes tedious. While our operations do combine views and optimize design, they do not at present allow for a final fine-tuning of, for instance, font sizes. Such global controls are not a part of our method, but would be highly helpful in any multi-view visualization design process.

Finally, our method is based on general principles in the sense of Kindlmann and Scheidegger \cite{kindlmann2014algebraic} and thus does not take into account an explicit task specification. While this focus was deliberate, since meaningfully characterizing user tasks is a significant challenge of its own that would also explode the design space, we still believe that exploring how different types of general user tasks could guide the evaluation of relationships and the presentation of operations is an important topic for future research.

% How more stuff could have been added

% How it was challenging to build this high-level functionality, and how
% it is merely a prototype to demonstrate the concept
% How some operations could be made more general

%-----------------------------------------------
\section{Conclusion}
We presented semantic snapping, a semi-automatic guided method that allows for incrementally refining multi-view visualizations. While previous work on multi-view visualizations has given us guidelines and constraints for reasoning about and improving visualizations, we further operationalized these concepts by (1) specifying relations between views precisely, and (2) proposing how each relation can be resolved by an operation. Each operation is a step in the semantic space with two axes representing the consistency and compactness. Furthermore, we presented a prototype implementation of our method, where users can perform operations to gradually refine a multi-view visualization design. In the future, believe that our approach to specifying relations and corresponding operations can be applied to more elements of multi-view visualizations such as their layout. Furthermore, many additional rules and guidelines for single visualizations could be adapted to or extended for multi-view visualizations.
%% if specified like this the section will be committed in review mode
\acknowledgments{
The research presented in this paper was supported by the MetaVis project (\#250133) funded by the Research Council of Norway as well as the VIDI project (\#813558) funded by the the Trond Mohn Foundation in Bergen, Norway.}

\bibliographystyle{abbrv-doi}

\bibliography{vis2021-semanticsnapping}

\end{document}

%% The ``\maketitle'' command must be the first command after the
%% ``\begin{document}'' command. It prepares and prints the title block.

%% the only exception to this rule is the \firstsection command
%\firstsection{Introduction}

\maketitle

This is an additional case study that could not be included in the paper due to space restrictions.
In this example, we illustrate the resolution of a \textit{hallucinator} and \textit{multiples} relation using the cars dataset. We begin with the three bar chart views arrayed on the canvas in Figure~\ref{semsnap:fig:example_3}.1. These views plot average horsepower against: number of cylinders (leftmost view), wheel drive (middle view) and aspiration (rightmost view). Each of these views constitute different pairwise groupings that share the same data domain plotted to the y-axis (average horsepower). This describes a \textit{multiples} relation. We furthermore see that all three views use the color channel to visualize the average price, but with different data domains. This constitutes a \textit{hallucinator}~(Figure~\ref{semsnap:fig:example_3}.1a). We resolve this \textit{hallucinator} with a \textit{homogenize} operation that changes the visual output to green~(Figure~\ref{semsnap:fig:example_3}.1b), in addition to homogenizing the data domains. This moves us from pos.~1 at the origin of our semantic space a more consistent position in the space~(pos. 2). 

With consistent color outputs across all views~(Figure~\ref{semsnap:fig:example_3}.2),
% It doesn't really "resolve" it since they are still multiples, only with the same domains, whereas integrate actually removes the relation
as a consequence of the \textit{multiples} relation between the three views, we are able to \textit{homogenize} the data domains of the three views~(Figure~\ref{semsnap:fig:example_3}.2a-b).
%we next choose to resolve the \textit{multiples} relation that we identified in our initial canvas by \textit{homogenizing} the data domains of the two rightmost views to match the leftmost
view~(Figure~\ref{semsnap:fig:example_3}.2a-b). 
Our model offers a sanity check for any explicit operation to \textit{homogenize} the data domain~(Figure~\ref{semsnap:fig:example_3}.2c). This is to help avoid a potential backwards operation in semantic space, i.e., create a less compact and/or less consistent visualization), that unreasonably distorts the data. However, the user ultimately validates whether the views they wish to combine are semantically the same. The option to undo their move is always available. With successful \textit{homogenization} of the y-axis data domains in all views we have arrived to pos.~3 in semantic space. Our resulting visualization is both more semantically consistent and compact than it began~(Figure~\ref{semsnap:fig:example_3}.3).